Magnetic Resonance in Medicine

# Accelerated and quantitative three-dimensional molecular MRI using a generative adversarial network

Jonah Weigand-Whittier[1] | Maria Sedykh[2] | Kai Herz[3,4]  | Jaume Coll-Font[1,5]  | Anna N. Foster[1,5]  | Elizabeth R. Gerstner[6] | Christopher Nguyen[1,5,7,8]  | Moritz Zaiss[2,3,9]  | Christian T. Farrar[1]  | Or Perlman[1,10,11]

**Correspondence**
Or Perlman, Department of Biomedical Engineering, Tel Aviv University, Tel Aviv 6997801, Israel.
Email: orperlman@tauex.tau.ac.il

**Funding information**
CERN openlab cloud computing grant; H2020 Marie Skłodowska-Curie Actions, Grant/Award Number: 836752 (OncoViroMRI); National Institutes of Health, Grant/Award Numbers: P41-RR14075, R01- CA203873, R01-EB03008

**Purpose:** To substantially shorten the acquisition time required for quantitative three-dimensional (3D) chemical exchange saturation transfer (CEST) and semisolid magnetization transfer (MT) imaging and allow for rapid chemical exchange parameter map reconstruction.

**Methods:** Three-dimensional CEST and MT magnetic resonance fingerprinting (MRF) datasets of L-arginine phantoms, whole-brains, and calf muscles from healthy volunteers, cancer patients, and cardiac patients were acquired using 3T clinical scanners at three different sites, using three different scanner models and coils. A saturation transfer-oriented generative adversarial network (GAN-ST) supervised framework was then designed and trained to learn the mapping from a reduced input data space to the quantitative exchange parameter space, while preserving perceptual and quantitative content.

**Results:** The GAN-ST 3D acquisition time was 42–52 s, 70% shorter than CEST-MRF. The quantitative reconstruction of the entire brain took 0.8 s. An excellent agreement was observed between the ground truth and GAN-based L-arginine concentration and pH values (Pearson's $r > 0.95$, ICC $> 0.88$, NRMSE $< 3\%$). GAN-ST images from a brain-tumor subject yielded a semi-solid volume fraction and exchange rate NRMSE of $3.8 \pm 1.3\%$ and $4.6 \pm 1.3\%$, respectively, and SSIM of $96.3 \pm 1.6\%$ and $95.0 \pm 2.4\%$, respectively. The mapping of the calf-muscle exchange parameters in a cardiac patient, yielded NRMSE $< 7\%$ and SSIM $> 94\%$ for the semi-solid exchange parameters. In regions with large susceptibility artifacts, GAN-ST has demonstrated improved performance and reduced noise compared to MRF.

**Conclusion:** GAN-ST can substantially reduce the acquisition time for quantitative semi-solid MT/CEST mapping, while retaining performance even when facing pathologies and scanner models that were not available during training.

Christian T. Farrar and Or Perlman contributed equally to this work.

For affiliations refer to page 1911

## 1 | INTRODUCTION

Semi-solid MT and chemical exchange saturation transfer (CEST) are MRI techniques that provide unique contrast, based on saturation transfer (ST). While semi-solid MT provides a means for studying macromolecules, lipids, and myelin,[1] CEST enables the detection of millimolar concentrations of mobile proteins, peptides, and metabolites.[2] However, most CEST imaging protocols provide a semi-quantitative contrast by way of the magnetization transfer ratio asymmetry (MTR$_{asym}$) analysis. While this metric has shown clear value in a variety of clinical studies,[3-5] it is dependent on numerous factors such as chemical exchange rate, volume fraction of the exchangeable solute protons, water longitudinal relaxation rate, radiofrequency (RF) saturation time, radiofrequency saturation power, and water transverse relaxation rate, all of which must be modeled across a variety of exchangeable proton pools.[4,6] Additionally, while clinical amide proton transfer weighted CEST protocols can now be acquired in just a few minutes,[7] the full Z-spectrum acquisition required for the analysis of multiple exchangeable proton pools (e.g., via Lorentzian model fitting) requires significantly longer acquisition times.[8] In addition, for quantification of the exchange parameters, the acquisition of multiple Z-spectra with different saturation powers is typically needed,[9] resulting in even longer acquisition times.

A variety of methods were previously developed for accelerating ST-weighted MRI.[10] Prominent examples include parallel imaging and transmission,[11] compressed sensing[12] ultrafast Z-spectroscopy,[13] and pulsed steady-state CEST sequences.[14] Recently, rapid developments in deep learning have been harnessed for semisolid MT/CEST imaging. Fully connected neural networks have been used to rapidly extract Lorentzian fitted parameters[15] and apparent exchange dependent-relaxation parameters from Z-spectra[16] and to predict 9.4T CEST contrast from 3T data.[17] Convolutional neural networks have been used for B$_0$ correction,[18] SNR enhancement,[19] tumor classification,[20] assessment of tumor progression,[21] and eightfold acceleration of CEST-weighted image acquisition.[22]

Magnetic resonance fingerprinting (MRF)[23] is a rapid and quantitative imaging paradigm that uses pseudo-random acquisition schedules to acquire unique signal trajectories, which are then matched to an existing database or "dictionary" of Bloch-equation-based simulated signals, providing a de facto pixelwise estimation of the underlying magnetic properties. Although initially developed for quantification of water T$_1$ and T$_2$ relaxation times, MRF has recently been modified and expanded for the quantification of semi-solid MT and CEST exchange parameters,[6,24,25] in an attempt to ameliorate the above-mentioned challenges of conventional ST-weighted imaging methods.

In many pathologies, the chemical exchange parameters of multiple proton pools (e.g. MT and amide) vary simultaneously and must be included in any MRF simulated dictionary. This leads to an exponential growth in dictionary size and hence very long dictionary generation and parameter matching times. The application of deep learning for accelerating water pool T$_1$ and T$_2$ MRF is an increasingly investigated field, where various approaches demonstrated a marked potential.[26-31] However, due to the larger number of tissue parameters that must be matched for CEST/MT fingerprinting, the implementation of neural networks for exchange parameter map reconstruction is more complicated than that employed for "conventional" T$_1$/T$_2$ MRF, and require a separate optimization and research endeavor.[32,33]

Accordingly, several deep-learning-based approaches have been developed specifically for shortening the reconstruction part of the semisolid MT/CEST MRF imaging pipeline.[34,35] While these strategies have demonstrated promising results, the acquisition time is still long and requires the acquisition of T$_1$, T$_2$, and B$_0$ maps, constituting an obstacle for routine clinical adoption.[32] Moreover, applying CEST-MRF for multislice imaging would further increase the scan time.

Recently, a unique machine learning approach has demonstrated the ability to learn the hidden and complicated relations (manifold) between two paired image categories[36] and generate an approximation of the appropriate image-pair for a given input. This approach, termed conditional generative adversarial network (GAN),[37] is built on two competing neural networks, a generator and a discriminator, which are trained simultaneously. The generator aims to synthesize convincingly realistic samples while the discriminator estimates the probability that a sample comes from the training data category.[38] During training, the generator gradually learns to create more convincing models based on the discriminator's feedback. The GAN framework is highly modular, and adversarial models have been shown to be effective for a variety of applications. In particular, the conditional GAN "pix2pix" architecture was developed for image-to-image translation problems.[37] Conditional GANs have shown great performance in translational tasks involving natural images, such as the synthesis of night views from pictures taken at daytime, generation of full object photos from edges, and maps from aerial photographs. The promise of conditional GANs has recently been translated and expanded into medical imaging, where this strategy was employed for cross modality synthesis (e.g., MRI to CT) and transformation between T$_1$ and T$_2$-weighted MRI maps.[39]

Here, we hypothesized, that a modified conditional GAN framework could be designed and trained to learn the manifold that links between raw semisolid MT/CEST-MRF encoded images and their quantitative exchange parameter image counterparts. Moreover, we assumed that an efficient quantification could still be obtained as the number of raw MRF encoded images is reduced, thereby allowing a substantial shortening of the acquisition time. Finally, to transform the developed approach into a clinically attractive and practical protocol, we have combined the MRF acquisition block with a three-dimensional (3D) Snapshot CEST readout module,[40] allowing rapid whole-brain (or any other organ) multi-slice imaging.

## 2 | METHODS

### 2.1 | ST-oriented GAN (GAN-ST) architecture

A supervised learning framework (Figure 1) was designed based on the conditional GAN architecture.[37] The generator was a U-Net convolutional network aiming to synthesize two proton exchange parameter maps (volume fraction and exchange rate), for either the semisolid MT or the CEST compound exchangeable protons. The discriminator aimed to predict whether the images are the "real" corresponding quantitative images, or a "fake" (generator synthesized maps). The ground truth was obtained by a dictionary-trained fully connected semisolid MT/CEST-MRF neural-network (Figure 1B) that received the full MRF acquisition schedule as input ($M = 30$ raw images). GAN-ST was trained to yield the same quantitative maps by receiving only a partial subset of $N = 9$ raw MRF encoding images as input. For the human brain imaging scenario, the water $T_1$, $T_2$, and $B_0$ may vary significantly in the WM/GM/tumor tissues. Thus, to improve the reliability and accuracy of the ground-truth for this case, the $T_1$, $T_2$, and $B_0$ maps were acquired separately, quantified, and given as an additional direct input to the reference ground truth MT/CEST-MRF network, as performed and described by Perlman et al.[34] Notably, these three maps were not given to GAN-ST. To preserve the perceptual and quantitative content of the original quantitative images while retaining spatial continuity and

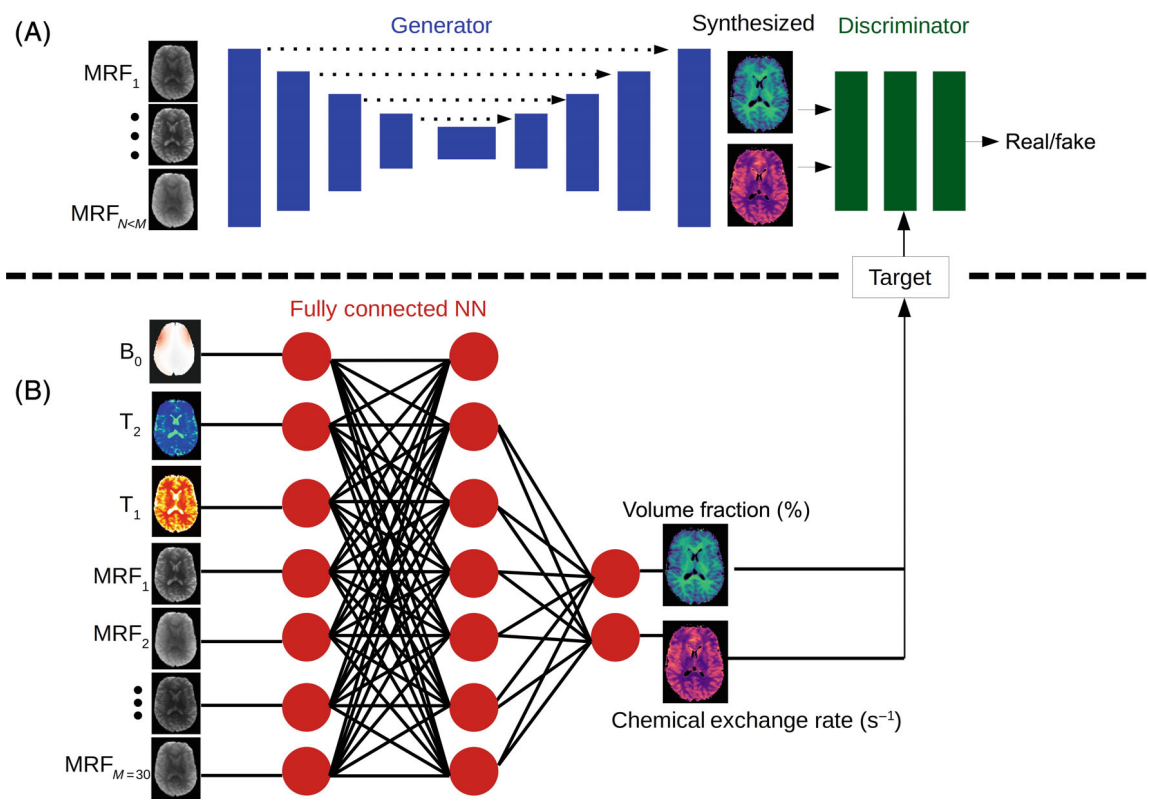

**FIGURE 1** Generative adversarial network (GAN)-saturation transfer (ST) architecture. (A) A conditional GAN framework, receives $N$ raw, molecular information encoding semi-solid magnetization transfer/chemical exchange saturation transfer images, and is trained to simultaneously output the quantitative proton volume fraction and the exchange rate maps. (B) A fully connected neural network, receiving the full-length raw magnetic resonance fingerprinting image series ($M > N$) pixelwise, as well as $T_1$, $T_2$, and $B_0$ maps, and yielding the reference proton volume fraction and exchange rate maps.[34] The output of this network was used for training GAN-ST.

**TABLE 1** Human imaging data cohort

| Imaging scenario | Properties | Training set | Validation Set | Test Set |
| --- | --- | --- | --- | --- |
| Brain | Number subjects | Five healthy volunteers | 2 GBM patients | 1 GBM patient and 1 healthy volunteer |
| Brain | Number images[a] | 611 | 166 | 164 (85 - patient, 79 - volunteer) |
| Brain | Scanner/coil/site | Prisma (64-channel, Boston) Prisma (64 channel, Tubingen) Skyra (64-channel, Boston) | Skyra (32 channel, Boston) | Prisma (64 channel, Erlangen) and Trio (32 channel, Erlangen) |
| Calf | Number subjects | 7 healthy volunteers | — | 1 cardiac patient |
| Calf | Number images[a] | 416 | — | 52 |
| Calf | Scanner/coil/site | Prisma/single element coil/Boston | — | Prisma/single element coil/Boston |

Abbreviation: GBM, glioblastoma.

[a]Sevenfold data augmentation was performed using translations and horizontal/vertical image flips. The numbers mentioned in the table represent the acquired images prior to augmentation.

smoothness in the GAN-ST output, the following total loss ($L_{total}$) function was used:

$$L_{total} = \lambda_1 L_1 + \lambda_2 L_{adv} + \lambda_3 L_{tv} + \lambda_4 L_p \quad (1)$$

Where $L_1$ is the pixelwise $l_1$ content loss, $L_{adv}$ is the adversarial loss determined by the discriminator, $L_{tv}$ is the total variation loss, and $L_p$ is the perceptual loss.[41] The latter was defined as the $l_2$ loss between the feature maps obtained from activating a pretrained deep convolutional network architecture (VGG19[42]) on the ground truth and on the GAN-ST approximated quantitative maps. $\lambda_{1,2,3,4}$ are the loss weights, determined using a separate validation image set (Table 1). The training was performed for 400 epochs with a batch size of 4. The learning rates of the generator and discriminator were 0.0001 and 0.0005, respectively. The method was realized in Keras[43] and implemented on a desktop computer equipped with a single Nvidia GeForce RTX 3080 GPU.

## 2.2 | CEST phantoms

A set of L-arginine (L-arg, chemical shift = 3 ppm) phantoms was prepared by dissolving L-arg (Sigma-Aldrich) in a pH 4 Buffer, at a concentration of 25, 50, or 100 mM. The phantoms were titrated to different pH levels between 4.0-6.0, and placed in a dedicated holder, inside a container filled with saline/PBS. The phantoms were 3D scanned twenty times at two different imaging sites (Tubingen and Boston) using two scanner models (Prisma and Skyra, Siemens Healthineers, respectively). At each scan, a different subset of 6–7 vials was used from the general range of 4–6 pH with an L-arg concentration of 25, 50, or 100 mM. The imaging was repeated after physically and randomly rotating/moving the phantom. The phantoms were independently prepared at each site.

## 2.3 | Human imaging subjects

All in vivo measurements were performed under approval by the local ethics/IRB committee. Each subject gave written, informed consent before the study. A total of 17 subjects were imaged and allocated into the training, validation, or test set, as described in Table 1. The subjects were scanned at three imaging sites (Tubingen, Boston, and Erlangen) and were either healthy volunteers, glioblastoma (GBM) patients, or cardiac patients. The separate validation set was used for setting the hyperparameters (e.g., the number of epochs, as determined by early stopping), and the same determined training parameters were used for all imaging scenarios (phantom, brain, and leg). The test set was designed to impose a challenging evaluation environment, aiming to explore the GAN-ST robustness and ability to extrapolate beyond the environment of the training samples. In particular, while all training subjects were healthy volunteers, the test set included a GBM patient, a cardiac patient, and a healthy volunteer imaged at a different site and scanner model, which were not used in the training set.

## 2.4 | MRI acquisition

The MRI experiments were conducted at three imaging sites using four 3T clinical scanners consisting of three different models (2×Prisma, Trio, and Skyra scanners,

Siemens Healthineers) and three coil types (64-channel head coil, 32 channel head coil, and a single-element leg coil). All acquisition schedules were implemented using the Pulseq prototyping framework[44] and the open-source Pulseq-CEST sequence standard.[45] The MRF protocol generated $M = 30$ raw, molecular information encoding images, using a spin lock saturation train ($13 \times 100$ ms, 50% duty-cycle), which varied the saturation pulse power between 0 and 4 $\mu$T (average pulse amplitude, the complete and exact saturation pulse train parameters are provided in the data availability statement).[34] The saturation pulse frequency offset was fixed at 3 ppm for L-arginine phantom imaging[6,33] or varied between 6 and 14 ppm for semisolid MT brain/leg imaging.[34] The saturation block was fused with the 3D centric reordered EPI readout module described by Mueler et al.[40] and Akbey et al.[46] providing a 1.8/1.8/2.5 mm isotropic resolution for phantom/whole-brain/calf-muscle imaging. The field of view was set to $256 \times 224 \times 156$ mm$^3$, echo time = 11ms, flip angle = 15°. The full 3D MRF acquisition ($M = 30$) took between 2:21 to 2:53 (min:s), depending on the scanner and coil configuration. For brain imaging, the same rapid readout module and hybrid pulseq-CEST framework were used for acquiring additional $B_0$, $T_1$ and $T_2$ maps, via WASABI,[47] saturation recovery, and multi-echo sequences, respectively, resulting in a total scan time of 8.5 min.

## 2.5 | Data analysis

### 2.5.1 | Phantom data preprocessing

In vitro images with no L-arginine vials, partial vials, or severe image artifacts were removed. The 239 remaining images were split into groups of 222 training images, from 20 phantom scans, and 17 test images from a different phantom. Canny edge detection and circle Hough Transforms were used for background masking and vial segmentation, respectively, implemented in Python. Sevenfold data augmentation was performed using translations and horizontal/vertical image flips.

### 2.5.2 | In vivo preprocessing

All images were motion-corrected and registered using elastix.[48] Gray-matter and white-matter segmentation was performed using statistical parameter mapping (SPM)[49] from a $T_1$ map. Quantitative reference CEST-MRF maps were obtained using a fully connected neural network trained on simulated dictionaries, where all $M = 30$ raw input measurements were taken as input. For brain imaging, pixelwise $T_1$, $T_2$, and $B_0$ values were also incorporated as direct inputs to the NN. For a detailed description of the CEST-MRF reconstruction and quantification procedure see the recent publication by Perlman et al.[34]

### 2.5.3 | Statistical analysis

Pearson's correlation coefficients were calculated using the open-source SciPy scientific computing library for Python.[50] Intraclass Correlation Coefficients (ICC) were calculated using the open-source Pingouin statistical package for Python.[51] The structural similarity index (SSIM)[52] was computed using the SSIM-python imaging library (PIL). In all box plots, the central horizontal lines represent median values, box limits represent upper (third) and lower (first) quartiles, whiskers represent $1.5 \times$ the interquartile range above and below the upper and lower quartiles, and circles represent outliers. Statistics in the text are presented as mean $\pm$ SD. Differences were considered significant at $p < 0.05$.

## 3 | RESULTS

### 3.1 | Phantom study - exchange parameter quantification performance

Representative GAN-ST generated exchange parameter maps are shown in Figure 2A,B. An excellent agreement between the GAN-ST generated and CEST-MRF-based L-arg concentration maps was observed, with an average normalized root mean-squared error (NRMSE) of 1.8 $\pm$ 0.1%, a SSIM of 0.975 $\pm$ 0.005 (Figure 3C,D), and a significant correlation across all slices (Pearson's $r = 0.967$, ICC = 0.844, $p < 0.0001$) (Figure 3A). GAN-ST generated, and CEST-MRF proton exchange rate maps were similarly correlated (Pearson's $r = 0.961$, ICC = 0.778 $p < 0.0001$, Figure 3B), with a NMRSE of 1.9 $\pm$ 0.1% and SSIM of 0.973 $\pm$ 0.005 (Figure 3C,D).

### 3.2 | Phantom study: direct estimation of concentration and pH

To explore the GAN-ST ability for direct estimation of the pH and compound concentration, we created another set of ground-truth reference images, where all pixels in each segmented vial were replaced by the pH-meter measured pH, and the analytic-scale determined L-arg concentration. GAN-ST was retrained while employing these images as the target, followed by an estimation of the pH and L-arg concentration in a different phantom test

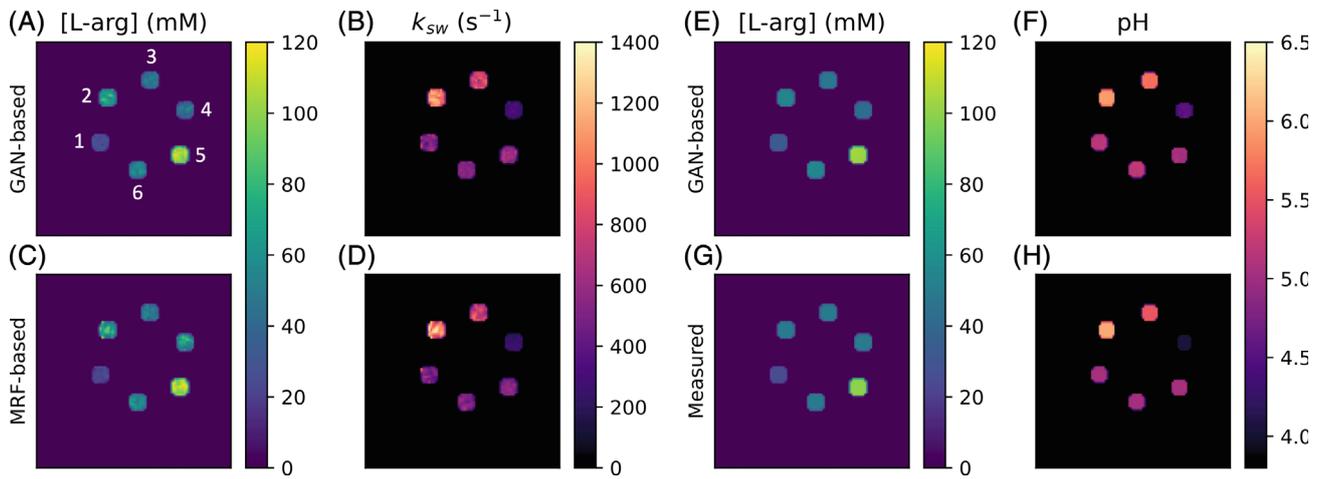

**FIGURE 2** Generative adversarial network (GAN)-saturation transfer (ST) in vitro image results. (A,B) L-arginine concentration (A) and exchange rate (B) maps from GAN-ST-based reconstruction, obtained with $N = 9$. Vials are numbered 1–6. (C,D) Full-length chemical exchange saturation transfer-magnetic resonance fingerprinting-based L-arginine concentration (C) and exchange rate (D) maps, obtained with $M = 30$. (E,F) GAN-ST-based ($N = 9$) concentration (E) and pH (F) maps. (G,H) Concentration (G) and pH (H) maps obtained using gold-standard non-MRI measures.

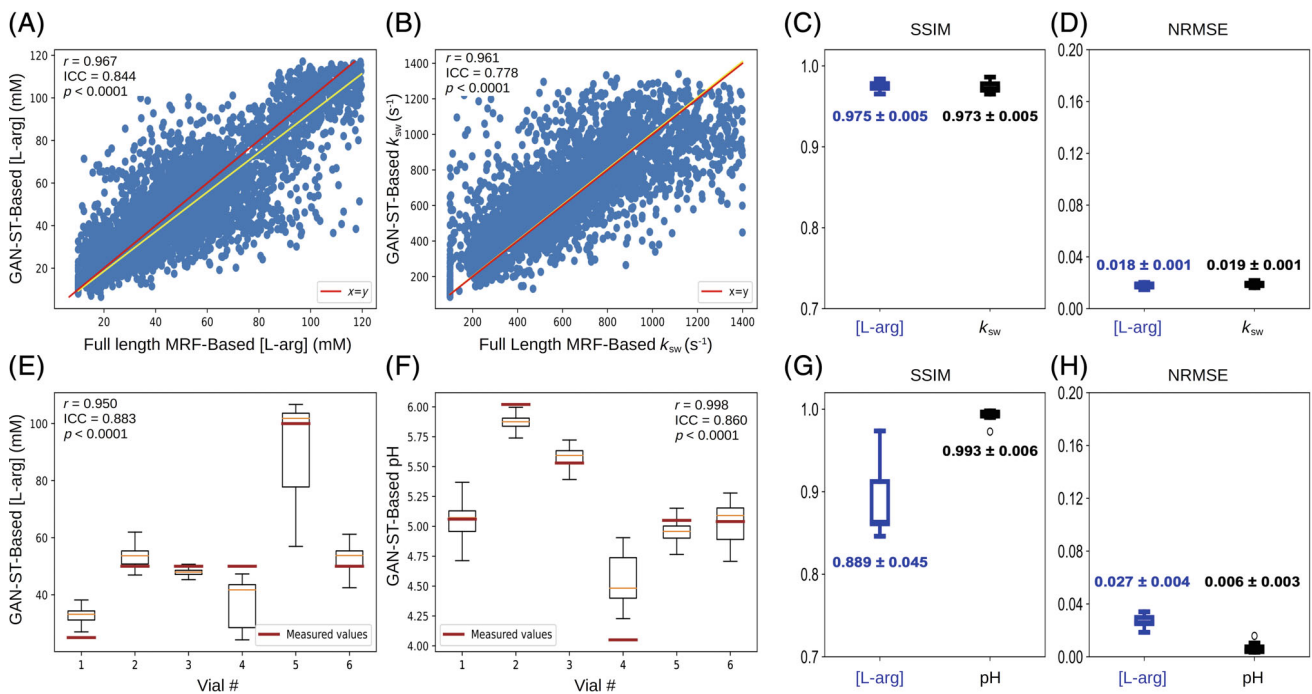

**FIGURE 3** Statistical analysis and quantitative assessment of Generative adversarial network (GAN)-saturation transfer (ST) performance in vitro. (A,B) Correlation between GAN-ST based and chemical exchange saturation transfer-magnetic resonance fingerprinting-based concentration (A) and exchange rate (B) maps across the entire three-dimensional volume of an L-arginine phantom. (E,F) Box plots showing the distribution of per-vial GAN-ST-based L-arg concentration (E) and pH (F) maps with measured values indicated. Vial numbers are based on Figure 2A. (C,D,G,H) Structural similarity index and normalized root mean squared error for concentration/exchange rate (C,D) and concentration/pH (G,H) maps.

set (Figure 2E–H). GAN-ST generated concentration maps were in good agreement with measured values, yielding an NRMSE of 2.7 ± 0.4% and SSIM of 0.889±0.045 (Figure 3G,H) and a significant correlation across all slices

(Pearson's $r = 0.950$, ICC = 0.883, $p < 0.0001$, Figure 3E). Similarly, GAN-ST generated pH maps were in good agreement with measured values, yielding an NRMSE of 0.6 ± 0.3% and SSIM of 0.993 ± 0.006 (Figure 3G,H) and a significant correlation across all slices (Pearson's $r = 0.998$, ICC = 0.860, $p < 0.0001$) (Figure 3F).

**FIGURE 4** Quantitative semisolid magnetization transfer (MT) parameter maps from a healthy volunteer, scanned at a site and scanner model that were not used during training. (A–D) Generative adversarial network (GAN)-saturation transfer (ST)-based semi-solid MT proton volume fraction maps, obtained with $N = 9$. (E-H) chemical exchange saturation transfer (CEST)-magnetic resonance fingerprinting (MRF)-based semisolid MT proton volume fraction maps, obtained with $M = 30$. (I-L) GAN-ST-based semi-solid MT proton exchange rate maps, obtained with $N = 9$. (M-P) CEST-MRF-based semi-solid MT proton exchange rate maps, obtained with $M = 30$. The red arrows indicate regions with susceptibility artifacts.

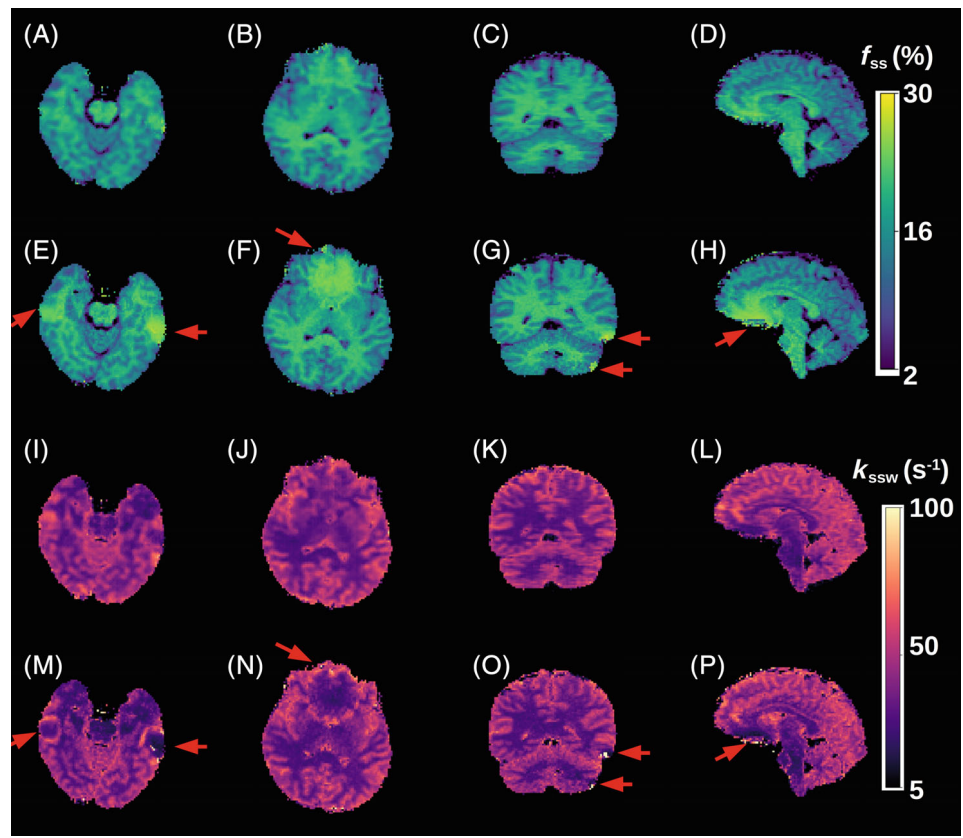

## 3.3 | In vivo study—brain parameter quantification

A comparison between the GAN-ST results with the number of raw molecular encoding images set to $N = 9$ and the full-length MRF-based reference ($M = 30$) for four representative slices is shown in Figure 4 (healthy volunteer scanned at a site and scanner model that were not available in the training cohort) and Figure 5 (tumor patient). The accelerated GAN-ST output was visually very similar to the CEST-MRF reference (average SSIM > 0.925, average NRMSE < 5.2%, Figure 6C,D,G,H). Moreover, in regions with large susceptibility artifacts, GAN-ST demonstrated improved performance and reduced noise compared to MRF (red arrows in Figures 4 and 5). Although the training cohort included only healthy volunteers, GAN-ST was able to output CEST-MRF comparable parameter maps, even in complex tumor and edema containing image slices (Figure 5). The resulting WM/GM $f_{ss}$ estimated by the GAN-ST approach for all nontumor-containing slices was $18.7 \pm 2.1 / 13.2 \pm 2.5\%$, compared to $18.7 \pm 2.0 / 12.4 \pm 2.7\%$ using the full-length MRF. The WM/GM $k_{ssw}$ estimated by the GAN-ST approach was

$36.2 \pm 6.1 / 51.6 \pm 8.5$ Hz, compared to $33.9 \pm 5.2 / 49.1 \pm 8.5$ Hz, by the full-length MRF reference, with significant correlation between individual pixel values obtained by both methods (Pearson's $r = 0.90$ and 0.75, ICC = 0.88 and 0.72, for $f_{ss}$ and $k_{ssw}$, respectively, $p < 0.001$, Figure 6A,B).

## 3.4 | In vivo study—calf-muscle parameter quantification

The GAN-ST-based in vivo calf exchange parameter quantification (Figure 7) was characterized by an SSIM > 0.94 and an average NRMSE < 7% (Figure 8). A significant correlation was observed between the GAN-predicted and reference CEST-MRF-based parameters, although the semi-solid volume fraction proton quantification was in better agreement with CEST-MRF than the exchange rate ($r = 0.73$, ICC = 0.71, $p < 0.001$, and $r = 0.51$, ICC = 0.45, $p < 0.001$, respectively).

## 3.5 | Acquisition, training, and inference times

GAN-ST was able to accelerate the scan time by 70% as it required the acquisition of only $N = 9$ raw molecular information encoding images instead of $M = 30$. This translated into a CEST/MT protocol acquisition time of only 42–52 s, depending on the scanner model hardware and the number of coil channels. Moreover, for the brain imaging scenario, GAN-ST circumvented the need to acquire separate $T_1$, $T_2$, and $B_0$ maps, providing a total acceleration of about 91% (44 s instead of 8.5 min). The total training time was 4.66/8.71/12.81 h, and the 3D inference time was 0.29/0.54/0.80 s, for the phantom, in vivo calf muscle, and in vivo brain, respectively.

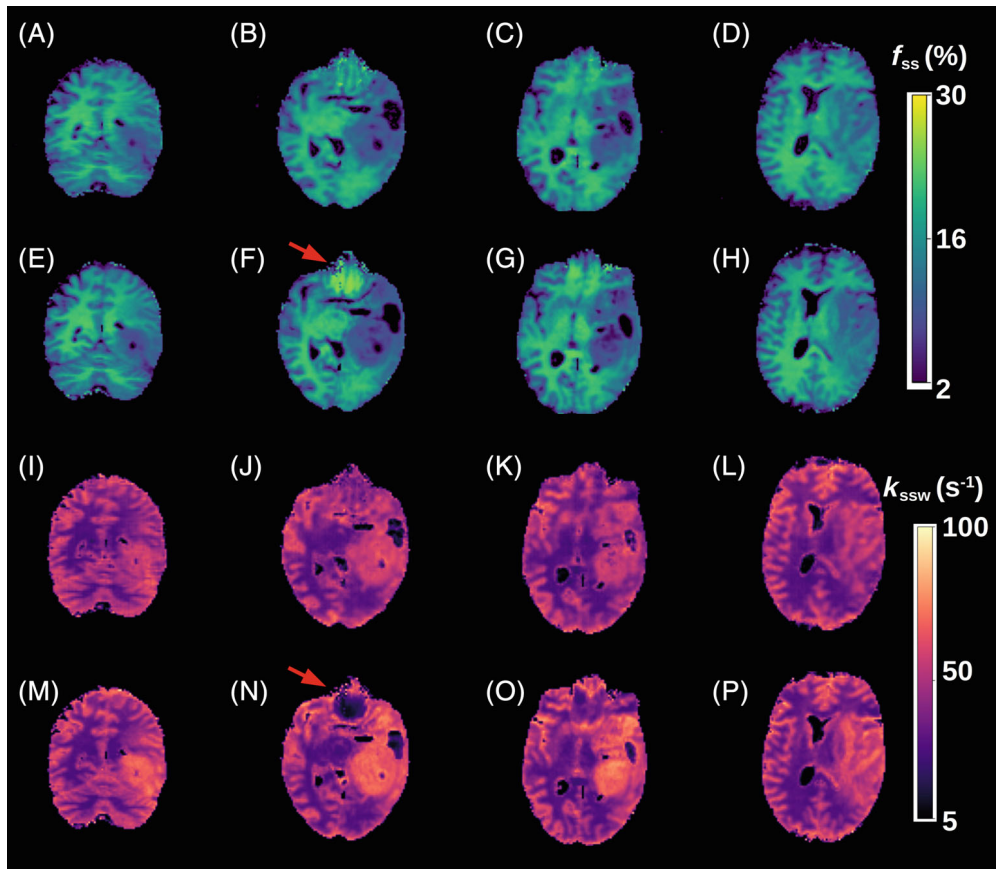

**FIGURE 5** Quantitative semisolid magnetization transfer (MT) parameter maps from a glioblastoma patient. (A–D) Generative adversarial network (GAN)-saturation transfer (ST)-based semi-solid MT proton volume fraction maps, obtained with $N = 9$. (E–H) chemical exchange saturation transfer (CEST)-magnetic resonance fingerprinting (MRF)-based semisolid MT proton volume fraction maps, obtained with $M = 30$. (I-L) GAN-ST-based semi-solid MT proton exchange rate maps, obtained with $N = 9$. (M-P) CEST-MRF-based semi-solid MT proton exchange rate maps, obtained with $M = 30$. The red arrows indicate regions with susceptibility artifacts.

## 4 | DISCUSSION

In recent years, the CEST contrast mechanism has been increasingly studied for a variety of medical applications.[5] The molecular information provided by CEST, most commonly by the amide proton transfer (APT) effect, has provided added clinical value compared to traditional $T_1/T_2$-weighted imaging. For example, endogenous CEST signals have been correlated with tumor lesion enhancement following Gadolinium injection.[4,53] Furthermore, CEST-weighted images were able to better discriminate treatment related changes from tumor progression.[54] MT signals stemming from semi-solid macromolecules have also been shown to be beneficial for cancer characterization and monitoring, either as a standalone method[55] or in combination with CEST.[56] In addition, MT has demonstrated potential as a biomarker for pathological skeletal muscle[57] and has long been known for its importance for multiple sclerosis imaging.[58] However, both CEST and semi-solid MT imaging are highly sensitive to the acquisition parameters used, as well as to changes in water pool relaxation. Moreover, these methods are prone to bias stemming from the analysis metric used and are subjected to contaminations from the signals originating from other tissue metabolites and compounds.[59] All the aforementioned challenges have motivated the development of quantitative approaches for semi-solid MT/CEST imaging.[60,61] In the context of clinical imaging, it is clearly important to also strive for short acquisition times. Semisolid MT/CEST MRF has recently been suggested as a rapid and quantitative molecular imaging pipeline.[32] However, as the multiprotocol conventional clinical routine is already lengthy (e.g., 30-40 min for each GBM patient monitoring session), it is essential to further accelerate semi-solid MT/CEST MRF, rendering it a cost-effective addition to clinical imaging protocols. The methodology proposed in this paper has made several contributions toward that cause. (1) The CEST-MRF saturation block was fused with a highly efficient and rapid snap-shot readout, allowing 3D semi-solid-MT or CEST-MRF acquisition in about 3 min, or 8.5 min (when $B_0$, $T_1$, and $T_2$ maps are also acquired based on the same 3D readout). (2) It employed GANs to further accelerate the 3D molecular scan time by reducing the required number of signal trajectory acquisitions by 70%, requiring acquisition times of less than 1 min. (3) The GAN-ST images have demonstrated the ability to mitigate the noise arising from field inhomogeneity and susceptibility artifacts (e.g., near the sinuses and the eyes, Figures 4 and 5). (4) The GAN-ST was able to extrapolate beyond the training data properties, as demonstrated using a subject scanned at a site and scanner model that were not used for training (Figures 4 and 6E–H).

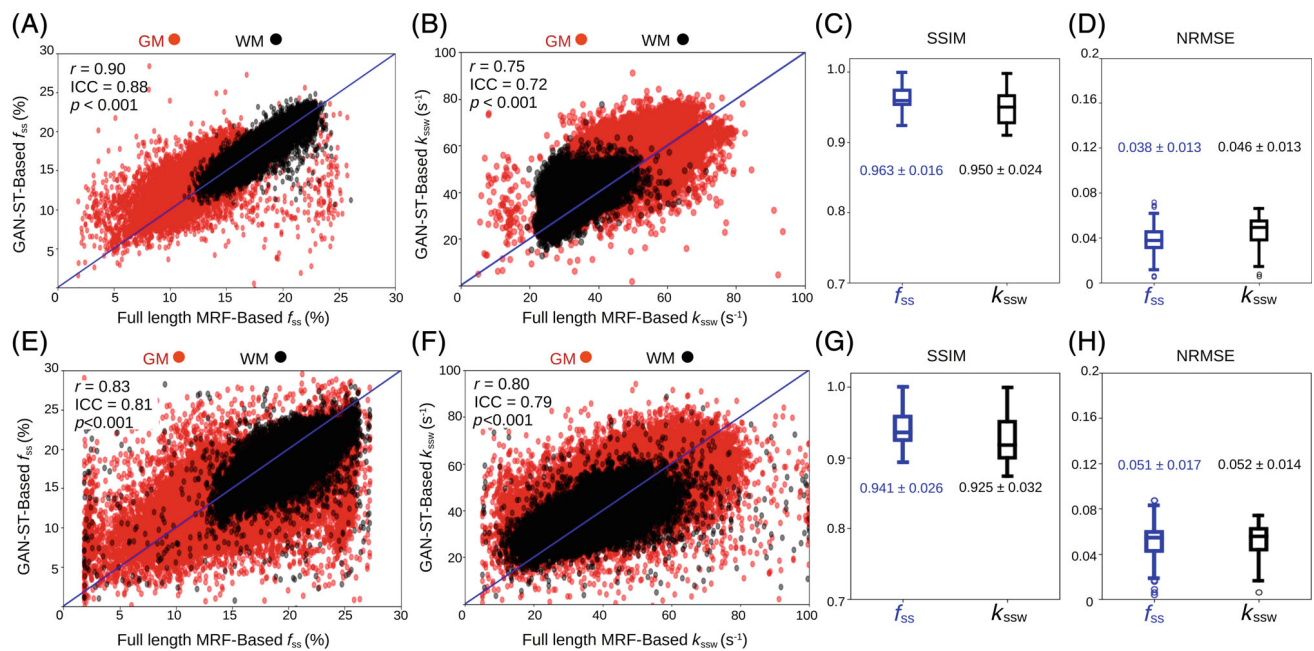

**FIGURE 6** Statistical analysis and quantitative assessment of the generative adversarial network (GAN)-saturation transfer (ST) performance in the in vivo brains of a tumor patient (A–D) and a healthy volunteer (E–H). (A,B) Correlation between all GAN-ST-based proton semi-solid magnetization transfer (MT) proton volume fractions (A) and exchange rates (B) for the entire brain in the WM/GM, and the corresponding pixel values obtained using chemical exchange saturation transfer (CEST)-magnetic resonance fingerprinting (MRF). Notably, the GAN-based $f_{ss}$ values in the WM are in better agreement with MRF refernce than the GM (Pearson's $r = 90$ compared to 0.74, respectively, $p < 0.001$), due to the myelin-rich content of the WM. (E,F) A similar analysis for the healthy human volunteer scanned at a site and scanner that were not available during training. (C,D,G,H) Structural similarity index metric and normalized root mean squared error for the tumor patient (C,D) and healthy volunteer (G,H).

In terms of the reconstruction (or parameter quantification from raw MRF images), one of the main limitations of "classical fingerprinting," is the lengthy dictionary matching times, where millions of simulated signals need to be compared to the experimental images (e.g., by means of a pixelwise dot-product matching). The framework presented here allows the reconstruction of quantitative proton volume fraction and exchange rate maps of the entire 3D volume acquired, in less than a second. Pixelwise matching using fully connected neural networks has previously been used for rapid semisolid MT/CEST MRF reconstruction.[34] However, the use of the GAN architecture, with its inherent spatial dependencies and

U-Net structure, has also provided the ability to mitigate susceptibility artifacts (Figures 4 and 5).

The semi-solid proton exchange rate was generally slower in the GAN-ST-based images, compared to CEST-MRF, in the tumor regions (Figure 5I–P), as well as in some anatomical parts of the calf (Figure 7I–K). This is in line with previous quantification attempts that faced similar challenges and reduced accuracy with the semisolid MT exchange rate.[35,62] These reports have attributed the poor discrimination ability for this property to the relatively small signal fluctuations that were observed as the exchange rate was varied. In this work, the reduced accuracy can also be explained by the intentionally challenging datasets design, where tumor data were not provided during the training. This assumption is supported by the increased correlation observed between the CEST-MRF and GAN-ST exchange rates for a healthy volunteer that was evaluated (Figure 4, and Figure 6F compared to Figure 6B). The phantom data quantitative exchange parameter maps obtained using GAN-ST with 70% acceleration were in excellent agreement with the CEST-MRF-based reference (Figures 2 and 3). Moreover, the same GAN architecture has allowed the direct estimation of the pH and compound concentration, measured

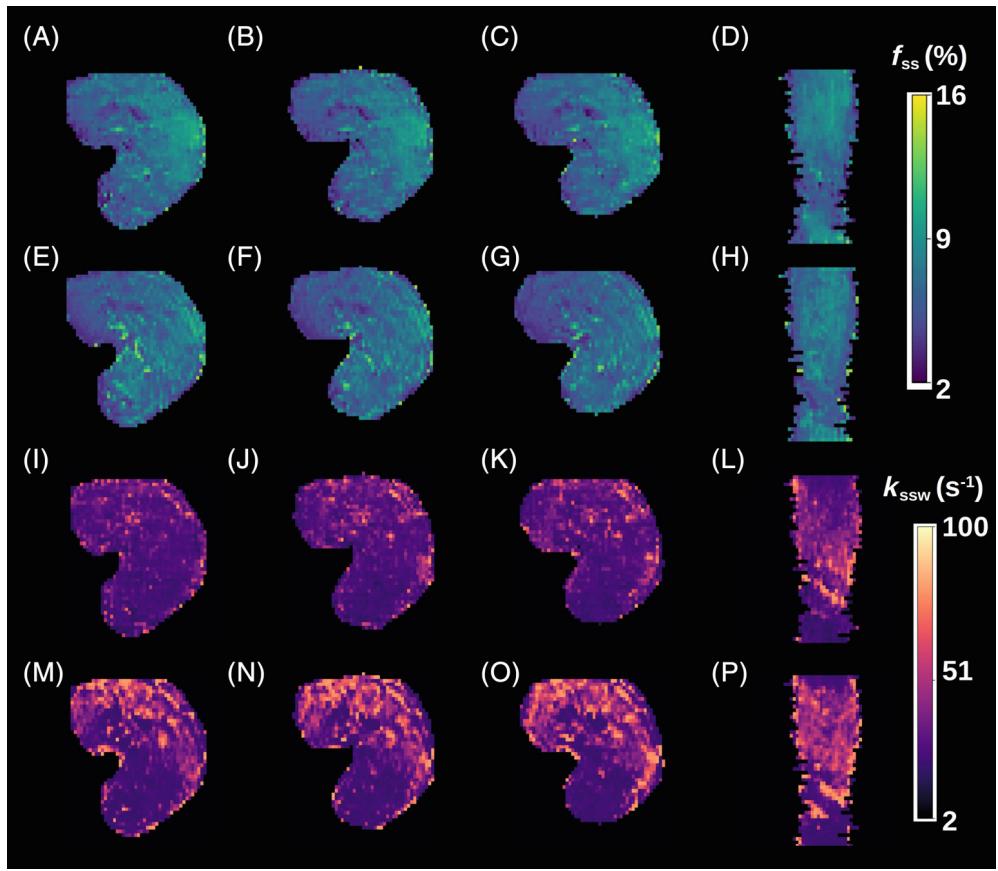

**FIGURE 7** Quantitative semi-solid magnetization transfer (MT) parameter maps from the calf muscle of a cardiac patient. (A–D) Generative adversarial network (GAN)-saturation transfer (ST)-based semi-solid MT proton volume fraction maps, obtained with $N = 9$. (E–H) chemical exchange saturation transfer (CEST)-magnetic resonance fingerprinting (MRF)-based semisolid MT proton volume fraction maps, obtained with $M = 30$. (I–L) GAN-ST-based semi-solid MT proton exchange rate maps, obtained with $N = 9$. (M–P) CEST-MRF-based semi-solid MT proton exchange rate maps, obtained with $M = 30$.

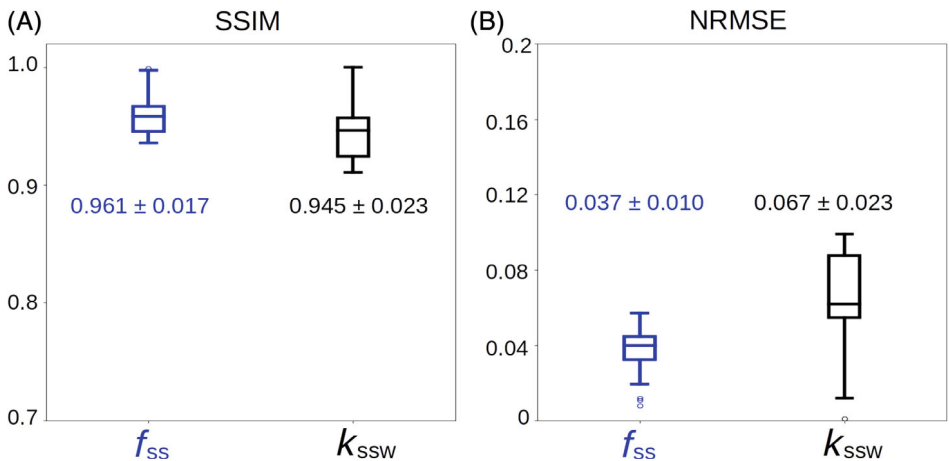

**FIGURE 8** Statistical analysis and quantitative assessment of the generative adversarial network (GAN)-saturation transfer (ST) performance in the calf-muscle of a cardiac patient. (A) Structural similarity index metric. (B) Normalized root mean squared error

using gold-standard non-MRI measures (pH-meter and analytical scale). The worst performance was obtained for the vial containing pH = 4 (Figure 3E,F). This is not surprising, given the base-catalyzed exchange rate of L-arginine, which results in small CEST signal amplitudes at low pH due to the slow exchange rate. While direct mapping from raw CEST data to pH is very attractive, future work should explore methods for obtaining in vivo ground-truth reference pH data.

There is a trade-off between the level of acceleration (manifested by the number of raw MRF images acquired,

$N$) and the parameter quantification performance. For example, increasing $N$ from 9 to 20, improves the GAN-ST visual similarity between the tumor semi-solid MT proton exchange rate, and those obtained by CEST-MRF (Figure S1). Similarly, the $k_{ssw}$ NRMSE decreased from $0.052 \pm 0.014$ using nine input images to $0.045 \pm 0.013$, using 20 input images. In addition, the correlation between the non-tumor tissue exchange parameters for GAN-ST compared to CEST-MRF is improved from $r = 0.90$, ICC $= 0.88$ ($p < 0.001$) and $r = 0.75$, ICC$=0.72$ ($p < 0.001$) for nine input images, to $r = 0.92$, ICC$=0.90$ ($p < 0.001$) and $r = 0.77$, ICC = 0.74 ($p < 0.001$) using 20 input images. While the quantitative maps obtained with a drastic acceleration of 70% are deemed both visually and quantitatively satisfactory, future expanded clinical evaluations could determine more accurately the degree of acceleration level sufficient for retaining a correct diagnosis.

This work demonstrates the feasibility of using a GAN-based framework to accelerate ST MRF. Notably, while the in vivo human studies were focused on semi-solid MT imaging, only in vitro experiments were conducted for CEST. The main reason is that a recent work has shown that accurate in vivo CEST-MRF requires a more complicated imaging strategy, which involves a serial acquisition of $T_1$, $T_2$, $B_0$, MT, and CEST data, and their integration in a sequential reconstruction pipeline.[34] In particular the CEST reconstruction pipeline receives the MT reconstructed parameter maps as input. While the current study serves as the first demonstration that GAN-based architectures can be used to accelerate CEST-MRF, future work should be performed to validate its application in-vivo, which will require integrating the GAN architecture in a more complicated reconstruction pipeline.

Several additional steps could improve the GAN-ST performance. First, the training set used here was intentionally composed of healthy volunteers only, aimed to examine the extrapolation ability of the method. In addition, although 3D acquisitions create a large number of images, the number of training subjects scanned was relatively small (Table 1). Significantly increasing the training cohort, and training on a variety of pathological cases, are expected to boost the accuracy of the method. Second, the performance of the proposed method is dependent on the original discrimination ability of the M-length acquisition schedule, as only images from the end of the acquisition schedule can be removed, and not the beginning, due to the acquired spin history induced by earlier acquisitions. Therefore, the parameter discrimination ability could be further improved, while benefiting from the spatial denoising and extrapolation capabilities demonstrated here, by combining the proposed GAN approach with an optimized acquisition schedule, which could be discovered using recently developed deep-learning-based sequence optimization approaches.[63,64]

## 5 | CONCLUSION

The GAN-ST framework has demonstrated the ability to accelerate 3D acquisitions of semisolid MT and CEST mapping by 70% while maintaining excellent agreement with full-length CEST-MRF-based reference maps and retaining performance across unseen pathologies and scanner models. Furthermore, GAN-ST has shown improvements over CEST-MRF in regions with large susceptibility artifacts. GAN-ST has exhibited promising initial results in direct estimation of compound concentration and pH from MRF encoded images.

## AFFILIATIONS

[1]Athinoula A. Martinos Center for Biomedical Imaging, Department of Radiology, Massachusetts General Hospital and Harvard Medical School, Charlestown, Massachusetts

[2]Institute of Neuroradiology, Friedrich-Alexander Universität Erlangen-Nürnberg (FAU), University Hospital Erlangen, Erlangen, Germany

[3]Magnetic Resonance Center, Max Planck Institute for Biological Cybernetics, Tübingen, Germany

[4]Department of Biomedical Magnetic Resonance, University of Tübingen, Tübingen, Germany

[5]Cardiovascular Research Center, Cardiology Division, Massachusetts General Hospital, Charlestown, Massachusetts

[6]Massachusetts General Hospital Cancer Center, Harvard Medical School, Boston, Massachusetts

[7]Health Science Technology, Harvard-MIT, Cambridge, Massachusetts

[8]Cardiovascular Innovation Research Center, Heart, Vascular, and Thoracic Institute, Cleveland Clinic, Cleveland, Ohio

[9]Department Artificial Intelligence in Biomedical Engineering, Friedrich-Alexander Universität Erlangen-Nürnberg, Erlangen, Germany

[10]Department of Biomedical Engineering, Tel Aviv University, Tel Aviv, Israel

[11]Sagol School of Neuroscience, Tel Aviv University, Tel Aviv, Israel

## ACKNOWLEDGMENTS

The authors thank Tony Stöcker and Rüdiger Stirnberg for their help with the 3D EPI readout. The work was supported by the US National Institutes of Health Grants R01-CA203873, R01-EB03008, and P41-RR14075. The research was supported by a CERN openlab cloud computing grant. This project has received funding from the European Union's Horizon 2020 research and innovation programme under the Marie Skłodowska-Curie grant agreement No 836752 (OncoViroMRI). This paper reflects only the author's view, and the European Research Executive Agency is not responsible for any use that may be made of the information it contains.

## DATA AVAILABILITY STATEMENT

The MRI acquisition schedule presaturation blocks used in this paper are available in the Pulseq-CEST open-source format[45] at https://github.com/kherz/pulseq-cest-library, under the folders MRF_CEST_Larginine_3T_002_13SL_DC50_2500ms and MRF_CEST_MT_3T_003_13SL_DC50_2500ms. All fully trained networks required to reproduce the GAN-based CEST/MT reconstruction presented in this work are available at https://doi.org/10.6084/m9.figshare.20346369, as well as a sample phantom dataset. The python code for creating and training GAN-ST is available at: https://github.com/jweigandwhittier/GAN-ST. The full 3D human data used in this work are not publicly available due to participant/patient privacy.

## ORCID

*Jonah Weigand-Whittier*  https://orcid.org/0000-0002-6647-9769
*Kai Herz*  https://orcid.org/0000-0002-7286-1454
*Jaume Coll-Font*  https://orcid.org/0000-0001-9341-6838
*Anna N. Foster*  https://orcid.org/0000-0002-1838-7237
*Christopher Nguyen*  https://orcid.org/0000-0003-1475-2329
*Moritz Zaiss*  https://orcid.org/0000-0001-9780-3616
*Christian T. Farrar*  https://orcid.org/0000-0001-6623-8220
*Or Perlman*  https://orcid.org/0000-0002-3566-569X

## TWITTER
*Or Perlman*  @PerlmanOr

## SUPPORTING INFORMATION

Additional supporting information may be found in the online version of the article at the publisher's website.

**Figure S1.** Comparing the quantitative semisolid MT parameter maps from a GBM patient using $N = 9$ raw input images (A–D, M–P), $N = 20$ input images (E-H, Q-T), and CEST-MRF reference (I–L, U–X).

**Figure S2.** A Monte Carlo simulation study comparing the quantification ability of GAN-ST trained on 9 or 30 inputs.

**Table S1.** Comparing Monte-Carlo based simulation of GAN-ST compound concentration quantification with 9 or 30 inputs.

**Table S2.** Comparing Monte-Carlo based simulation of GAN proton exchange rate quantification with 9 or 30 inputs.

**Figure S3.** Comparing the quantitative parameter maps obtained using GAN-ST with $N = 9$ raw input images (bottom row) and the respective output from CEST-MRF with $N = 9$ raw input images (center row), for the phantom data presented in Figure 2. Clearly, reducing the number of input images from $N = 30$ (top row) to $N = 9$ severely degraded the quantification performance for CEST-MRF, whereas GAN-ST resulted in better agreement with the ground-truth and the full length CEST-MRF.